# The Rosetta Stone and Levels of Principled Inference to the Experience of Another Mind.

Kallum Robinson[1, 2, 3, φ], Giulio Tononi[2], Naotsugu Tsuchiya[3, 4, 5, 6, 7], Matteo Grasso[2, 4]

[1.] Berlin School of Mind and Brain, Faculty of Philosophy, Humboldt-Universität zu Berlin, Berlin, Germany.

[2.] Department of Psychiatry, University of Wisconsin-Madison, Madison, U.S.A.

[3.] School of Psychological Sciences, Faculty of Medicine, Nursing and Health Science, Monash University, Melbourne, Australia.

[4.] Advanced Telecommunications Research Computational Neuroscience Laboratories, Kyoto, Japan.

[5.] Turner Institute for Brain and Mental Health, Faculty of Medicine, Nursing and Health Science, Melbourne, Australia.

[6.] Center for Information and Neural Networks (CiNet), National Institute of Information and Communications Technology (NICT), Osaka, Japan.

[7.] Theoretical Sciences Visiting Program (TSVP), Okinawa Institute of Science and Technology Graduate University, Onna, Japan.

[φ] Correspondence to: kallum.robinson@monash.edu

# Abstract

The classical problem of *Other Minds* has dogged philosophers for millennia; asking if we have any way to truly understand the experience of another mind. We know our own intrinsic experience by *acquaintance*, but can only ever hope to possess an extrinsic *description* of another's, with the two separated by an *acquaintance gap*. Structural approaches aim to characterise experience in terms of a mathematical structure, and promise a 'Rosetta Stone'; that is, a principled method to translate the contents of experience into a mathematical structure. In this chapter, we examine two such approaches - the Qualia Structure Paradigm (Qstr) and Integrated Information Theory (IIT) - to ask what, if anything, they allow us to infer about another mind should they possess the Rosetta Stone they seek. Qstr proceeds inter-phenomenally; aiming to exhaustively characterise an experience by its relations to all other experiences and providing a necessary condition on the sameness of experience. IIT proceeds intra-phenomenally; conjecturing that a single experience is accounted for by the cause-effect structure unfolded from a substrate in a state, an *explanatory identity* that is both necessary and sufficient for the sameness of experience. While neither approach can cross the *acquaintance gap* to another's mind, they provide a common formal medium through which minds may be compared and in turn reduce the size of this gap. We identify the strength of inference with the levels of structural correspondence in Category Theory. These descend from isomorphism through strong and then weak adjunction, to a principled limit where inference runs out entirely. We conclude that these structural approaches and the existence of their respective Rosetta Stones would not solve the problem of *Other Minds*, but instead provide a system of principled constraints on what we may infer, which is far more than what has been justifiable before.

# 1. Introductions and the Problem of Other Minds

Can we hope to *truly* understand what another person experiences? Even accurately expressing the contents of our *own* conscious experience is a task of monumental difficulty (Nagel, 1974). These qualitative contents of experience, also known as qualia,[1] are precisely '*what it is like*' to be oneself in a given moment. Even if we struggle to fully describe them, we certainly do know our own qualia *intrinsically* (i.e., from our first-person perspective). Therefore, knowing another's qualia is inherently fraught, as we necessarily observe them from an extrinsic perspective (i.e., third-person relative to the experiencer). This complication is the classical problem of *other minds,* which has occupied philosophers since Descartes. Extrinsically, we may observe what another system[2] does and measure the mechanisms that make it do so, without thereby learning what, if anything, it is like to *be* that system (Nagel, 1974; Jackson, 1982, 1984).

In response to the difficulty of capturing qualia from any perspective, consciousness science has undergone a 'structuralist' turn in an effort to better characterise them via mathematical structures and spaces (Tononi, 2008; Balduzzi and Tononi, 2009; Fink et al., 2021; Lyre, 2022; Kleiner, 2024).[3] This chapter focuses on two such approaches. The first is the Qualia Structure paradigm (QStr), which characterises a quale through its web of relations to other qualia, presently by constructing relational structures from subjective similarity judgements (Tsuchiya & Saigo, 2021; Tsuchiya, 2025). It has been applied to the study of colour qualia (Zeleznikow-Johnston et al., 2023; Kawakita et al., 2025; Togashi et al., 2026), visual motion (Robinson et al., 2025), and object and facial identities (Rowe et al., 2025). The second is Integrated Information Theory (IIT), which, by contrast, begins by characterising the properties of experience through introspection and then asks what kind of substrate can account for them in physical terms (Albantakis et al., 2023; Hendren et al., 2024; Tononi and Boly, 2025). IIT has already been applied to accounts of why space (Haun and Tononi, 2019) and time (Comolatti et al., 2025) feel the way they do, and there is ongoing work focusing on the experience of objects (Grasso et al., in preparation). These accounts are introduced in chapter 18, and the complementarity of IIT and the QStr is discussed in chapter 11. Here, we use these two chapters as background to focus on a narrower question: what, if anything, do these structural approaches allow us to infer about *another* mind?

---

[1] Any moment of conscious experience, which we differentiate as a quale in the broad sense, is itself composed of many qualia in the narrow sense (Balduzzi and Tononi, 2009; Kanai and Tsuchiya, 2012; Lee-Youngzie et al., 2026). Sitting by the water in the afternoon sun is a broader quale, where the warmness of the sun or the sounds of the water are themselves narrower qualia.

[2] We use 'system' to denote any physical substrate that supports or specifies a conscious experience (a 'mind'). A system could in principle be a population of neurons, transistors on a silicon chip and so forth.

[3] Mathematical structures, such as metric spaces, offer a flexible framework for capturing both informational content and dynamics. Metric spaces consist of a set of points whose pairwise relationships satisfy specific conditions (depending on the nature of the space) and allow for a geometric interpretation of whatever resulting structure is formed by those points. For example, in the case of Qstr dissimilarity structures, individual points are qualia (such as individual colours) whose pairwise relationships reflect the dissimilarity ratings given by participants. For approaches that don't consider points as qualia, see (Tsuchiya et al., 2025). These structures are intended to better capture the quality of experience over and above more traditional binary approaches. Typically, binary judgements such as seen/unseen have been relied on to maximise the strength of inference in classic neural correlate of consciousness (NCC) research, however, they are unable to tell us what *it is like* to experience a stimulus.

## 2. Understanding the Need for a Rosetta Stone

Before outlining how structural approaches can be used to address the problem of *Other Minds*, it helps to draw a distinction here owing to Russell (1912), between knowledge by *description* — the sort we gain from a third-person specification— and knowledge by *acquaintance*— the sort of first-hand knowledge we gain by having an experience itself (Russell, 1912). In Strawson's words: "*the having is the knowing*" (Strawson, 2003). *Descriptions* such as mathematical structures are necessarily constructed from outside experience, and here there is good reason to think that the '*what it is like'*-ness is lost (Russell, 1927; Nagel, 1974; Jackson, 1982, 1984; Lee, 2024).

That said, if we possess a principled method by which we can translate our own *acquainted* experience into a *description* (such as a mathematical structure), then we have a means to infer *what it is like* to have the experience the structure corresponds to. Pursuit of a principled method to translate back and forth between structures and the experiences connected to them has been referred to elsewhere as the need for a 'Rosetta Stone' (Chalmers, 2012, 2023). Historically, the Rosetta stone was used to translate an unknown language (the Hieroglyphics of Ancient Egypt) into the known language of Ancient Greek by way of two identical blocks of text. Following the translation of the Rosetta Stone, historians' understanding of Hieroglyphics was so fundamentally transformed that they could use this newfound knowledge to read other blocks of Hieroglyphic text.

Applied to experience then, this metaphorical *Rosetta Stone* considers an *acquainted* experience to be the known block of text and the other unknown side as its corresponding *description* in terms of a mathematical structure. A principled method by which to translate any experience into a corresponding mathematical structure would 'solve' this problem, as is the broad aim of both Qstr and IIT. Like the real stone did for Hieroglyphics, a Rosetta Stone would provide us a means to account for the felt qualities of experience (qualia). With a Rosetta stone in hand, then in the same way that translation between two natural languages is possible because each language can express common meaning (i.e., point to the same referent in the world around us without *being* the thing referred to), two different experiences may in principle be compared by way of the common formal medium (mathematical structures) that correspond to them; returning us a possible solution to the problem of *other minds*.

With a Rosetta stone, such a solution to *Other Minds* would deliver a complete 'dictionary' of experience between us, in that each felt quality of our respective experiences can be translated into each other's one-to-one. If "*the having is the knowing*", this dictionary is therefore epistemically bound by acquaintance. In principle it could be formed between two systems having an identical experience, yet such a case is an exceptionally unlikely one, as we will discuss in §5. Below this upper bound a structure (however completely it captures experience) cannot deliver on the felt intrinsic character it characterises for another system, leaving us in turn with an *acquaintance gap.*

We argue that the size of this gap is not fixed and in fact leaves a great deal of principled inferences within reach. Structural alignment cannot transmit acquaintance with another subject's experience, but it can nevertheless constrain epistemically the range of phenomenal attributions warranted by third-person evidence. If we lay and stare at a starry sky together, I cannot know exactly what it is like to be you in that moment. Clearly though, my own acquaintance with darkness and the specks of light within it affords me at least a partial understanding. In other words, the structure corresponding to my own experience can serve as an anchor from which I may bring into alignment the structures of yours. The more our experiences share, the more our structures align and the more warranted my inferences will be about your experience—whether I share it or not.

The strength of inference across minds can be thought of on a gradient from strong to weak, determined by the degree to which our own experiential structures align with those of others. We identify this gradient with the levels of correspondence between structures in mathematical category theory (Tsuchiya et al., 2023), contingent on the hypothesis that both qualia structures and cause-effect structures (from IIT) are themselves categories, in the category theory sense (Tsuchiya et al., 2016; Tsuchiya, 2025). In principle, the strongest inference (an *isomorphism*) between structures of experience would be possible in cases where two minds have identical experiences. Below this isolated case, we argue inferences between structures of narrow qualia primarily take the form of category theory's *adjunction*: a coherent, systematic relationship between differing structures that are permissive of a lossy translation between them. We differentiate between *strong* and *weak* adjunction, depending on how constrained the inference is.[4] At their best, *strong adjunctions* deliver an imperfect dictionary—a translation that lands us near the meaning of another's experience, though not exactly. At the other end, *weak adjunctions* deliver the *grammar* of another structure. Grammar does not tell us what every word means, but how words are organised and what roles they can play. Likewise, a grammar of experience would not deliver another's contents, only their *kind*—that a substructure is organised as, say, one specifying a temporal experience rather than a spatial one. Finally, where structures of experience do not align at all, in that we can not anchor another structure to our own, we reach an inferential limit where nothing principled remains to be said.

First, we will briefly introduce the essential machinery of Qstr and IIT in §3 and §4, respectively, before discussing the descending levels of constrained inference that these approaches allow us to make about others' experiences in §5 and present efforts to validate each approach in §6 before concluding in §7.

[4] We use the terms *strong* and *weak* adjunction in an informal, non-category-theoretical sense, which makes no official discrete distinction in the strength of adjunction. The formal correlate of this spectrum is how far the adjunctions unit and co-unit (the maps comparing each round trip with the identity) depart from isomorphisms.

# 3. The Qualia Structure Paradigm's Rosetta Stone

The Qualia Structure paradigm aims to characterise a quale by its relations to all other qualia (Tsuchiya and Saigo, 2021; Tsuchiya, 2025). Across the realm of colour experiences for example, the narrower quale of red's *redness* is characterised partly by its more similar relation to the *orangeness* of orange than the *blueness* of blue. Empirically, these relations can be measured using similarity; by asking people to express how similar two experiences are from one another with a bounded numerical value and storing this information in a dissimilarity matrix. From this matrix, one can recover an underlying dissimilarity structure rendered in geometric space (Zeleznikow-Johnston et al., 2023; Kawakita et al., 2025; Robinson et al., 2025) which is an approximation of the putative full qualia structure. While similarity is used most often in practice, all manner of relations between qualia can be evaluated in principle. Other relations between experiences have been discussed elsewhere, such as 'association' (Smith et al., in prep) and 'preference' (Matsuyanagi et al., 2025) (Moul et al., in prep). Additionally, 'inclusion' (Lee-Youngzie et al., 2026) characterises how narrow and broad qualia are related within an experience.[5]

We understand Qstr here as offering an *inter-phenomenal Rosetta Stone* from the side of experience (Figure 1). This is because the structure recovered is not just an extrinsic property that happens to covary with experience, rather it is constructed from experience itself by a reporting subject acquainted with their own experience. These relations are *inter-phenomenal* in that they compare one experience to another. Because these relations are supplied from an *acquainted* experience, they constrain the *relata* they hold between. A sufficiently rich web of relations would (theoretically) fix the mapping between an experience and its structure one-to-one, in that an experience is exhaustively described by its relations to all other conceivable experiences in both broader and narrower senses.[6] It is important to be precise that this experience-structure pairing is epistemic in that it lies in the *provenance* of the structures (that its relations were supplied by a subject we *assume* to be conscious) and not in any intrinsic property of the mathematical object itself. Nothing within a dissimilarity matrix records what it actually *is*, and handed this without the labels of an assumed conscious subject, we may be unable to say what experience actually grounds it in the first place (Chapter 11, Box 1: *A Shared Ontology?*). For example, dissimilarity structures recovered from Large Language Models (LLMs) align extremely well with those recovered from humans (Kawakita et al., 2024), yet there is sufficient reason to doubt that these LLMs possess qualia (Findlay et al., 2025) and are instead reproducing the statistical regularity of the data they are trained on rather than possessing a principled method to translate experience into a mathematical structure.

In using these structures to address the problem of *Other Minds*, any alignment of Qualia Structures is necessarily based on such assumptions. I can compare my *redness* to my *orangeness* because I am acquainted with both experiences and then store a quantified aspect of this relation in a dissimilarity matrix. If we want to compare my *redness* to yours, one option is to compare the structures themselves. Indeed, dissimilarity structures have been compared without assuming shared labels using unsupervised Gromov-Wasserstein optimal transport (GWOT) both between groups (Kawakita et al., 2025) and between individual subjects (Togashi et al., 2026) (Chapter 3). In these cases, successful alignment shows that two structures share a form, that is, the relations between the points of the structure are highly similar. Correspondence between qualia structures is at most argued to be a necessary condition for any 'sameness' in experience, but not a sufficient one (Tsuchiya, 2025). In §5 we will return to how much 'sameness' such alignment licenses us to infer about another mind.

[5] Narrow *redness* is a constituent relation in broader qualia such as the experience of watching a sunset and therefore has further relations among all possible experiences that must be captured in the qualia structure (Lee-Youngzie et al., 2026).

[6] A guiding intuition of the paradigm is the Yoneda lemma from category theory, which holds that an object is determined, up to isomorphism, by the totality of its relations to the other objects of its category (Chapters 1 & 14) (Tsuchiya and Saigo, 2021; Tsuchiya et al., 2022). Applied to experience, the thought is that a quale may be characterised by the totality of its relations to all other qualia. Critically, an arrow of this kind (similarity, inclusion, association and so forth) can only be drawn by a subject acquainted with both of its relata. Your qualia are theorised to form a category to which you alone supply the relations (only you can say how similar your red is to your orange).

# 4. Integrated Information Theory's Rosetta Stone

Where many structural approaches (Rosenthal, 2010; Fink et al., 2021; Kob, 2023; Fleming and Shea, 2024; Kleiner, 2024; Tsuchiya, 2025) begin with the relations among reported (extrinsic) experiences and works towards their physical basis (Maier and Tsuchiya, 2026), IIT begins from the essential properties of experience (axioms) themselves and works to identify a substrate whose causal powers can account for these properties in physical terms (postulates) (Albantakis et al., 2023). These postulates are requirements in terms of cause-effect power that a physical substrate must meet in order to account for experience. A substrate that satisfies these postulates unfolds into a cause-effect structure - the full set of distinctions and relations that the substrate specifies about itself. IIT conjectures that the quality of an experience (the *what it is like*-ness) is entirely accounted for by the *form* of this cause-effect structure,[7] accounting for both the essential and 'accidental' properties of experience. These 'accidental' properties are those that give experience its specific content, such as colour, sound, space, time and so forth. This *explanatory identity* holds that every property of an experience should be accounted for by a corresponding property of the cause-effect structure (for a more comprehensive understanding, see chapters 11 & 18). If correct, this identity is the very *Rosetta Stone* outlined in §2, validated *intra-phenomenally* within one experience and between the cause-effect structure that accounts for it, providing a principled method to translate any experience into its corresponding mathematical structure (Figure 1). The cause-effect structure is *intraphenomenal* in that it exhaustively translates a single, momentary and immediate experience (*this* one) into its corresponding structure.

Where Qstr recovers its structure extrinsically from similarity judgments, IIT builds intrinsicality into its axioms before any question of an external observer arises. This difference in the provenance of the respective mathematical structures sets up a division of labour for the following section (§5). Alignment between structures recovered from reports (Qstr) can, at most, be a necessary (not sufficient) condition for any sameness in experience (§3). IIT's explanatory identity supplies sufficiency; if the quality of experience is fully accounted for by the form of the cause-effect structure, then two systems/substrates specifying identical cause-effect structures cannot differ in what their experiences are like.

[7] The cause-effect structure is interchangeably referred to in literature on IIT as both a Φ-structure and an intrinsic causal structure.

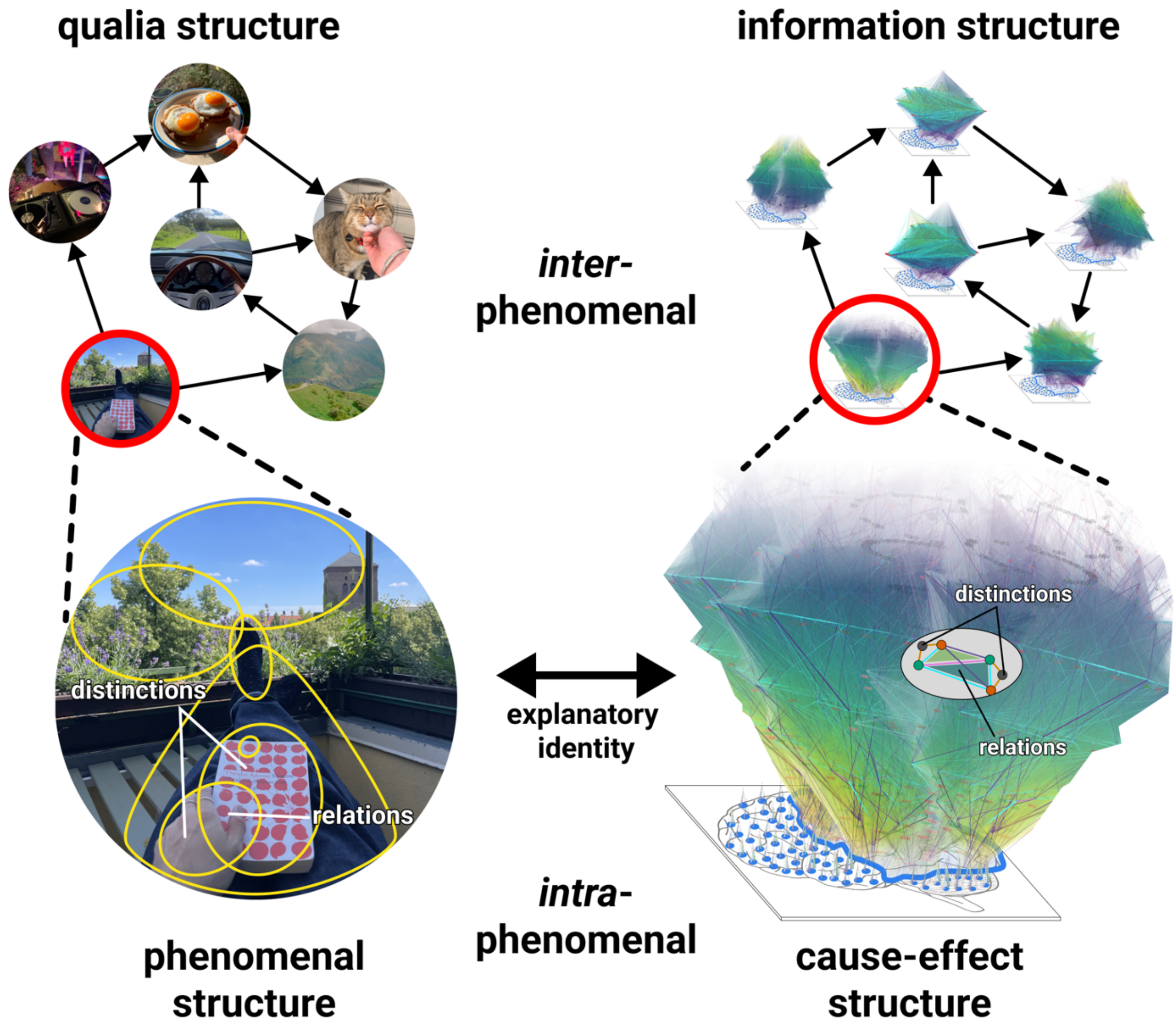

**Figure 1. Difference in intra- and inter-phenomenal approaches to the Rosetta Stone.**
Broadly speaking, both Qstr and IIT seek a Rosetta stone, that is, a principled method by which to translate experience (*acquaintance*) into a mathematical structure (a *description*), yet do so from different directions. Qualia structure captures the relations between individual experiences (*inter*-phenomenal). These structures capture the relations between broad qualia such as petting a cat or driving a car (shown), but also narrower qualia such as *this* red and *this* blue (Kawakita et al., 2025). Theoretically, a web of relations rich enough could constrain any individual quale to such a degree that it is exhaustively described by these relata. Qstr also aims to find correspondences between the qualia structure and *qualia-relevant structures from physical substrates* (summarised here as 'information structures') (Tsuchiya, 2025). While Qstr can accommodate cause-effect structures (or substructures thereof) from IIT in this aim (shown), in principle Qstr is theory-agnostic so long as they are specific enough to specify causal structures (Tsuchiya, 2025). IIT however, captures *intra*-phenomenal structure in that the corresponding cause-effect structure is an explanatory identity for immediate phenomenal experience (*this* experience) with nothing left over. IIT aims to do so by identifying the essential properties of experience and then formulating them in physical terms (in terms of the causal power of a substrate). IIT conjectures that the resulting unfolded cause-effect structure from this substrate can account for the phenomenal distinctions and relations composing the phenomenal structure, capturing both the essential and accidental properties of experience.

## 5. Levels of Inference to Another Mind

If correct, QStr and IIT allow us to characterise our own experience in systematic, structural terms (i.e., construct a Rosetta Stone). So how could we use them to say something about another's experience? As laid out in §2, the answer depends on how closely another mind's corresponding mathematical structure aligns with structures already shown to capture the contents of our *own* experience. In category theory, a structure of this kind is theorised to be a category, composed of; *objects* as points in the structure and *arrows* are directed relations between them, each running from one object to another and composing according to certain conditions. A correspondence between two such structures is then given by a *functor* - a rule carrying every object and arrow of one category to a counterpart in the other while preserving how they compose—much the same way a translation maps words and grammatical relations between languages. A *functor* is thus itself an *arrow*, now between whole structures rather than within one and the levels of inference we will outline in the following sections classify functors by how much they preserve on a "round trip" from one structure and back again. Round-trip mappings that return every arrow exactly as it was are most tightly constrained, while those that can only return us to a region of possible beginnings are loosely constrained.

The strongest level of inference for our purposes is an *isomorphism*, possible when two structures have a bi-directional mapping that preserves all object and arrow relationships, allowing any object or arrow to return to its exact starting point after a round trip. As the mapping between structures in category theory is concerned only with relations, it is possible to preserve relational structure yet change the meaning of the constitutive objects (see ch. 6).[8,9] Below isomorphisms sit *adjunctions*, which make up the vast majority of possible inferences between structures. This is because structures are almost certain to differ in varying degrees, and any translation between them is correspondingly lossy. Each relation that two structures share is one less way their corresponding experiences can therefore differ. As the number of shared structural relations accumulate, the possible ways that these experiences differ proportionally decreases, until the structures themselves leave no room for experience to differ at all. Given this, we differentiate between *strong* and *weak* adjunction, depending on this degree of structural similarity.

---

[8] A natural consequence of an isomorphic relationship is the inverted qualia thought experiment. Consider this classic case: if we were to construct a relational structure for colour experience, we could in principle invert it, preserving all the structural relations among colour experiences while transforming *redness* into *greenness,* leaving the inversion undetectable from the structure alone. However, this inversion would remain undetectable only as long as narrow *redness* is considered within other narrow qualia. Because Qstr considers narrow qualia in relation to all other qualia, including those broader qualia and those in other domains, it is highly unlikely such inversion would be undetectable. The same consideration applies to any content whose structure is asymmetric. See Chapter 6 for a full treatment of this question.

[9] Closely below isomorphisms are categorical equivalences. For our purposes, the correspondences expressed by categorical equivalences and isomorphisms can be treated as largely similar. A categorical equivalence is a round-trip mapping that returns not necessarily the original object itself, but an object isomorphic to it. In this qualified sense, it collapses only distinctions between objects that were already categorically indistinguishable up to isomorphism within their own category. Consider this example: assume the English words *sofa* and *couch* bear exactly the same relations to all other English words, i.e., they mean *exactly* the same thing. In translating these English words to French, one mapping sends both words to the French *canapé*, and the reverse map returns *canapé* to *couch*. Considering the round trip from *sofa* to *canapé* then back to *couch*, nothing is lost but the label, since no relation separated *sofa* from *couch* to begin with. Category theory does not register such structureless differences, and for our purposes categorical equivalence therefore behaves exactly as isomorphism does.

Why should inference scale with structural correspondence in this way? These levels are a consequence of characterising experience relationally. Recall that a quale is thought to be characterised by the totality of its relations to all other qualia (§2). Taken seriously, this conjectures that the identity of any narrow quale is fixed by all of its relations within the structure of the broad quale of which it is a constituent. Mapping between structures preserves this identity only if it preserves the relations, which is precisely what category theory is poised to accomplish. In what follows we descend this gradient, pairing at each level an example from language with an example from experience.

## 5.1 Isomorphism

An isomorphism is the strongest standard correspondence between two distinct categorical structures. It consists of structure-preserving mappings in both directions that are exact inverses of one another. If we begin with any object in category A, map it to category B, and then map it back to category A, the round trip returns exactly the original object. The same must hold when starting from any object in category B. Crucially, this exact round-trip recovery must also hold for every arrow in both categories, while preserving their sources, targets, identities, and composition. Between minds, such a mapping would constitute a complete *dictionary* —a method by which I could translate your qualia into my qualia and gain complete understanding of your experience in both broadest and narrowest senses.

To illustrate this, imagine we are both looking at my pet dog, Ned. IIT conjectures this experience is entirely explained by the cause-effect structures unfolded from our substrates in the state corresponding to the experiences we are having (Albantakis et al., 2023). Qstr conjectures that this experience can be described by our inter-phenomenal relations to all other experiences. We can both establish such structures for our own experience (§§3-4), so with a common mathematical formalism in place, the only way I could be completely sure that a specific part of *your* structure corresponds to looking at Ned would be for it to be *exactly* the same as the part I have already validated in mine.

However, if our structures align perfectly, then you and I would thus need to have *exactly the same experience* (Figure 2A). This means that when you look at Ned, you must have the same inter-phenomenal relations both between narrower qualia (the colour of his fur, the shape of his features) and broader qualia (what he represents conceptually, the memories associated with him). Clearly then, for you to have *my* memories of *my* pet dog, for you to feel the same emotions as me in the same way, you would, in effect, have to become my *self*.[10] In principle this closes the acquaintance gap since we would both know by acquaintance an experience with identical structures. Isomorphisms therefore, are possible only in theory, for it is hard to conceive of a way in which we could be having identical experiences in the broadest sense (being the same self). In practice two systems will almost certainly differ in some of their relations, changing the structure and therefore diverging the mapping at once. To handle these far more realistic cases, we may descend to *Adjunction.*

[10] It is worth noting that we would not have to be the same *physical* system. Small changes in neuron dynamics and interconnectivity could exist between us, since on IIT cause-effect structures are multiply realisable at the macro scale, where micro-level differences may be essentially washed out in selecting a configuration of intrinsic units that maximises cause-effect power (Marshall et al., 2018, 2024).

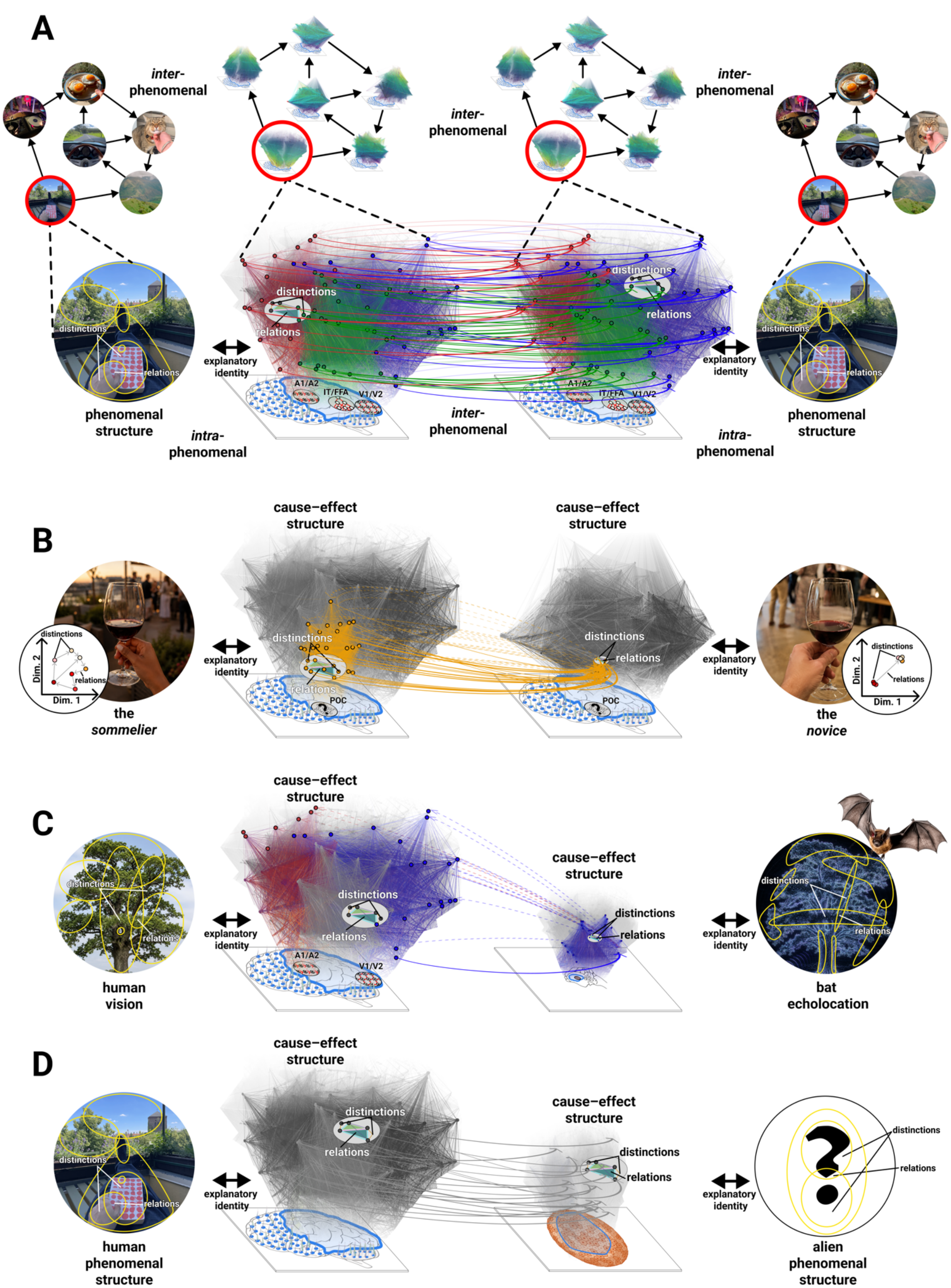
A
inter-
phenomenal
inter-
phenomenal
inter-
phenomenal
distinctions
relations
distinctions
relations
distinctions
relations
distinctions
relations
explanatory
identity
explanatory
identity
A1/A2
IT/FFA
V1/V2
A1/A2
IT/FFA
V1/V2
phenomenal
structure
phenomenal
structure
intra-
phenomenal
inter-
phenomenal
intra-
phenomenal
B
cause–effect
structure
cause–effect
structure
distinctions
relations
Dim. 2
Dim. 1
distinctions
relations
POC
distinctions
relations
POC
distinctions
relations
Dim. 2
Dim. 1
explanatory
identity
explanatory
identity
the
sommelier
the
novice
C
cause–effect
structure
cause–effect
structure
distinctions
relations
distinctions
relations
distinctions
relations
distinctions
relations
explanatory
identity
explanatory
identity
A1/A2
V1/V2
human
vision
bat
echolocation
D
cause–effect
structure
cause–effect
structure
distinctions
relations
distinctions
relations
distinctions
relations
distinctions
relations
explanatory
identity
explanatory
identity
human
phenomenal
structure
alien
phenomenal
structure

**Figure 2. Levels of Principled Inference in Comparing Experiential Structures.**

A, B, C & D correspondingly descend the gradient of inference outlined in §5. Inter-phenomenal structure reproduced from Fig.1 and shown in A theoretically extends to all subsequent cases in B, C & D. While it may be difficult to obtain a qualia structure from a bat or an alien given the unintelligibility of their reports, in principle and if they are indeed conscious, they would have such a structure in that their experiences can be related to one another. Regarding the colour coding of cause-effect structures shown throughout, IIT starts from introspection to characterise the distinctions and relations (the structure) that defines the phenomenal content of experience. IIT then formulates those phenomenal properties in terms of cause-effect power and attempts to find a corresponding substrate that reflects these same distinctions and relations. Each kind of content is conjectured to be accounted for by a kind of causal substructure unfolded from a kind of substrate constituting the main complex (highlighted in various colours). *Undirected 2D grids*—reminiscent of the connectivity of early visual cortex (V1/V2) and plausibly somatosensory cortex—should unfold into $\Phi$-folds accounting for the defining properties of spatial extensions (blue) (Haun and Tononi, 2019); *directed 1D grids*—reminiscent of the connectivity of early auditory cortex (A1/A2) and plausibly other areas contributing temporal contents (such as MT/V5 for visual motion)—should unfold into cause-effect substructures ($\Phi$-folds) accounting for the defining properties of temporal flow (red) (Comolatti et al., 2025); *rooted trees*—reminiscent of the connectivity patterns in the ventral stream of visual cortex (and culminating in IT/FFA) and likely other areas containing invariant mechanisms—should unfold into $\Phi$-folds accounting for the defining properties of conceptual hierarchy (green). **A.** An isomorphic relationship (§5.1) between cause-effect structures exists only when two systems have exactly the same experience. In this case alone a dictionary is formed where contents and structure can be directly translated across all modalities (returning solid arrows) **B.** A strong adjunct relationship (§5.2.1) is possible between different experiences that share common substrate features. The sommelier (left) has a certain phenomenal structure (represented by the theoretical qualia structure taken from Fig. 3) with a corresponding cause-effect substructure (orange) unfolded from a plausible physical substrate in the primary olfactory cortex (POC). Similarly, the novice (right) has a respective cause-effect substructure unfolded from their respective POC when they both taste the same wine. In aligning the sommelier's CES to the novice, the sommelier's richer structure maps down to the novice but does not precisely return (dotted arrows). By possessing commonalities in their respective structures, the sommelier is able to make a principled inference that the novice is indeed tasting wine (and not, say, beer or coffee), but cannot precisely say what it would be like to not possess a richer phenomenal structure. **C.** A weak adjunct relationship (§5.2.2) between my experience (left) of seeing a tree and a bat (right) seeing the same tree by echolocation. In mapping my unfolded cause-effect substructure specified by undirected 2D grids specified in V1/V2 (blue), I may identify that a bat has some *kind* of space experience, but can say nothing of *what it is like* to experience spatial extendedness by echolocation. For each distinction and relation in the bat's CES, there are many possible alignments in my own structure (spread of returning dotted arrows). In fact, some cause-effect substructures in the bat may even align better with substructures specified by A1/A2 (red), suggesting 'seeing' with echolocation for a bat could feel somewhat like 'hearing' for me. **D.** Failure to align given a lack of relations (§5.3) between my experience (left) and whatever *it is like* for another physical system that satisfies the postulates of IIT (right). In this case, while I may try to compare my CES to this alien system, nothing aligns (grey arrows with no return), so I am in no position to say what, at all, its experience is like.

## 5.2 Adjunction

An adjunction (or adjunct relationship) consists of a pair of functors that can be asymmetric, such that passing from one structure and back again does not return you to your starting object and arrow. An asymmetric mapping yields the best available approximation that the two structures themselves fix.[11] How good this approximation is depends on how much structure the two sides already share. Adjunction is thus best thought of on a spectrum, which in turn constrains the boundaries of the inferences we may make. At its strongest, the round-trip constrains our inferences to a narrow region, and the adjunction returns an 'imperfect dictionary'. At its weakest, the round-trip hardly constrains the inference at all, leaving us a grammar that permits inference to the *kind* of experience.

[11] Consider perhaps a concrete example between the whole number 3 and the real number 3.4 (as two categories ≤Z and ≤R, consider ≤ as arrows and integers or reals as objects) (Tsuchiya et al., 2023). The whole numbers fit inside the reals (this is one mapping), and any real number can be rounded down into the nearest whole number (this is another mapping *back*). Importantly all relations ≤ are kept in this mapping. Rounding 3.4 down to 3 and then reading 3 back as a real number (3.0) does not recover the original 3.4, but neither does it return an arbitrary number. It returns the closest available approximation under that mapping rule. This best-approximation relationship is exact, systematic and is therefore predictable. This kind of relationship is an adjunction.

### 5.2.1 Strong Adjunction

In a strong adjunction, the two structures share enough relations that the round trip, though lossy, is bounded within a narrow domain. For example, consider that Russian speakers are obliged to distinguish *Голубо́й/golubój* (light blue) from *синий/sínij* (dark blue)[12] when referring to a colour that is similar to purple and different to yellow. English offers the single word *blue* which covers both shades (Winawer et al., 2007). A Russian–English dictionary sends both *Голубо́й* and *синий* to the single English word *blue*, and with them the dissimilarity that held between them, which the mapping collapses to nothing.[13] The mapping (a translation) collapses the two words into one, and translating *blue* back into Russian cannot recover which of the two blues was originally meant. This is a *strong* inference by adjunction as the translation is lossy, but the loss is bounded and principled in that one remains within the domain of blue and not any other colour or, indeed, any other word.

What, then, does this kind of relationship look like when comparing our experiential structures? First consider for example, the qualia structure of a sommelier and a novice tasting the same seven wines (Figure 3). The sommelier likely possesses a richly differentiated qualia structure, both in the narrower sense (they can differentiate a wider range of notes for a single wine) and in a broader sense (they can differentiate a wider range of wines from one another; Fig. 3B). On the other hand, the novice's structure likely collapses these qualia into a smaller point, such that it is hard to separate one quale (a single note or broader taste of one wine) from another (Fig. 3A). If we were to align the two, the sommelier's fine distinctions would map down onto that narrow domain (Fig. 3C), yet the reverse mapping (from the novice to the sommelier), would return only a vague region within which the sommelier's experience must lie (Fig. 3D). This is not to say that the sommelier thereby has "acquaintance" with the novice's experience (nor conversely). Yet we can say something principled about the relationship between these structures: it is coherent in that we can map the sommelier "down" onto the novice in a principled manner, but lossy in that we cannot precisely return back to the sommelier in the same way as *Голубо́й* and *синий* collapse into *blue*. This remains though, an inter-phenomenal correspondence - locating the other's experience within a shared domain without necessarily certifying that the domain is genuinely shared in what is felt (§3).

---

[12] Pronounced as *guh-loo-boy* and *see-nee-y,* respectively.

[13] Strictly, an adjunction holds between *categories*, so for the purposes of this example, each language (English and Russian) must be construed as one. Additionally, we take each to be an *enriched category*, whose objects are words and whose hom-objects are the graded similarities among them (Tsuchiya et al., 2022). Enriching the category is necessary because similarity cannot serve as an arrow in an ordinary category because composition fail: for example, A may feel similar to B, and B to C, without A being similar to C. In an enriched category the composition law is instead an inequality on these graded values, which similarity can satisfy. A dictionary between Russian and English is thus a functor between enriched categories, carrying not only words to their counterparts but also the *similarities* among them to the similarities among those counterparts; thus obtaining the adjunction.

The same strong adjunction can be run intra-phenomenally. Following the method of IIT, I may unfold a cause-effect structure that entirely explains my current experience and parcel out which cause-effect substructures specify which contents - spatial extendedness from interconnected 2D undirected grids (Haun and Tononi, 2019) the flow of time from 1D directed grids (Comolatti et al., 2025) and, for our sommelier, the taste of wine from a substructure unfolded from a plausible substrate such as the primary olfactory cortex (Figure 2B). Both sommelier and novice could unfold such a substructure and the sommeliers should map down onto the novice just as their qualia structures did. Because a cause-effect structure is not a correlate of content but the content itself (by IIT's *explanatory identity*), alignment in substructures certifies that the two are tasting the same kind of thing (wine, and not say, coffee or beer). Indeed, on IIT's account of meaning as structure, this is the condition that ordinary communication between two people already depends on ((Zaeemzadeh and Tononi, 2024; Box 1). The lossy mapping still withholds from the sommelier what the wine actually *is like* for the novice, since that quale is set within the novice's own broad experience (Balduzzi and Tononi, 2009; Kanai and Tsuchiya, 2012; Lee-Youngzie et al., 2026). Strong adjunction thus lands us, either from Qstr or IIT, in a bounded region of the same domain near to but not identical with one's own experience.

### 5.2.2 Weak Adjunction

In a weak adjunction, by contrast, the two structures share so few relations that the round trip returns not a region within a domain, but at best the domain itself. Comparing structures at this level would be like encountering a word of a language that no living person speaks and for which no dictionary exists, and then attempting to translate it into English. This is not a hypothetical question, the Meroitic language is an extinct language of present-day Sudan, in use from c.300BC to c.400AD. The meanings of many Meroitic words are not known. However, by analysing the relational structure of a word (how it relates to other words in a sentence), linguists have been able to recover individual words' *kind;* i.e., they have recovered part of the *grammar* of the language (Rilly and De Voogt, 2012). This sort of analysis closely mirrors an adjunction between poset categories,[14] by bringing the role-structure so recovered within Meroitic into alignment with the role-structures of languages we already understand, which licenses attributions of *kind* (this word behaves as a determinant) but not of content. This same division will come to hold between minds.

[14] Part of working out Meroitic grammar relied on a back and forth pairing between words and surrounding words (Rilly and De Voogt, 2012). From a group of words, one can ask which positions (relative to other words) do all of these words appear in. From a group of positions, one can ask what words appear in all of these positions. These mirrored questions are an adjunction (more specifically, a Galois connection see: (Ganter and Wille, 1999)). Stable groupings that settled out from this back-and-forth mapping revealed which words behaved alike, and recurring elements such as the ending *-l(i)* (the determinant) could then be identified as grammatical morphemes.

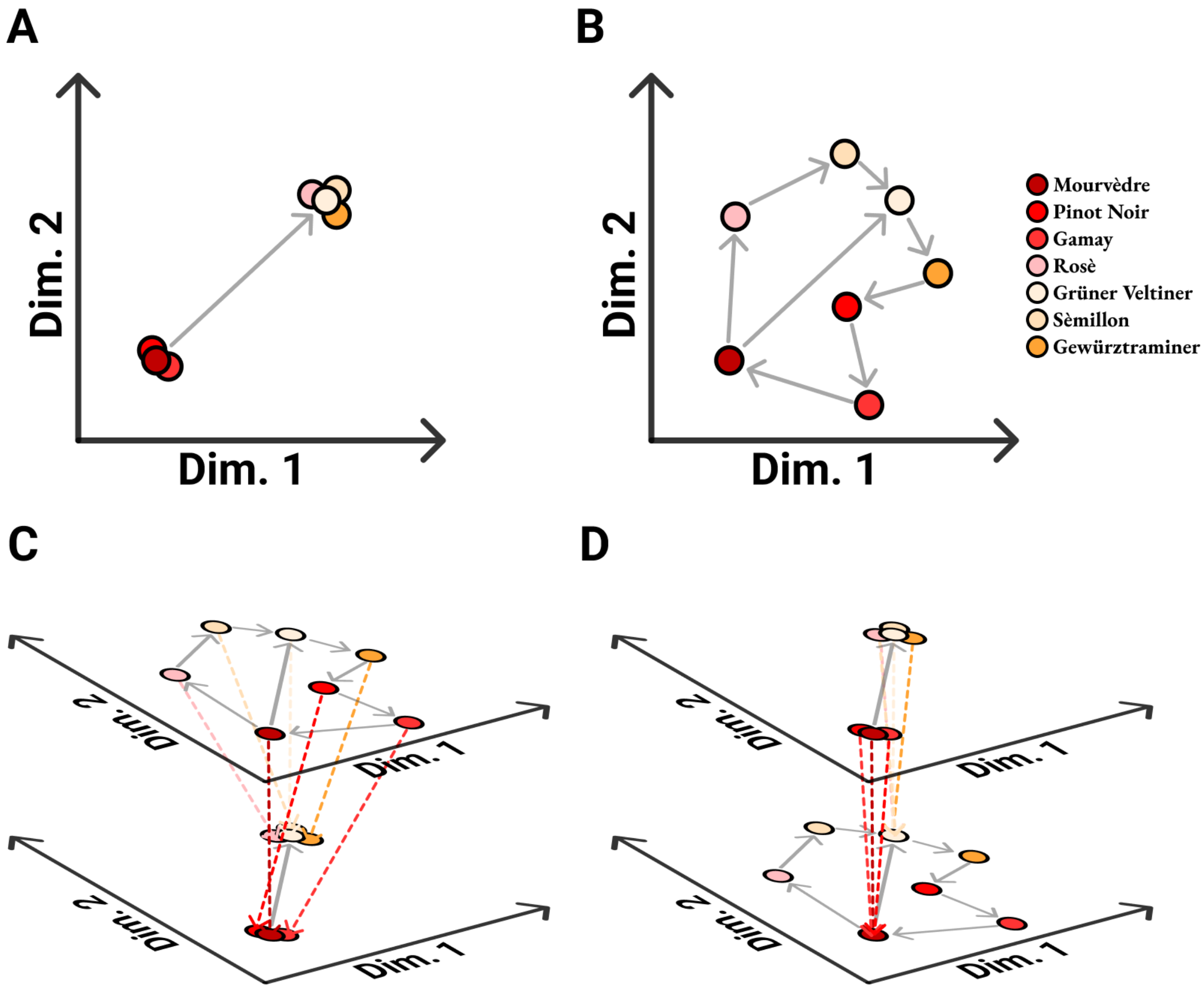

**Figure 3. Theoretical difference in structure between novice and sommelier's experience by adjunction.**

Plots in **A** and **B** represent the difference in the structure of experience of 7 wines (circles) in two dimensions of experience (Dim. 1 & Dim. 2) between the Novice (**A**) and the Sommelier (**B**). Each coloured circle represents the quale of tasting a different wine, represented by the shared key on the far right. The two dimensions of experience on the plot can be thought of as the two by which the greatest variation exists between the wines present. For example, Dim. 1 could be sweetness and Dim. 2 could be acidity. Arrows between wines reflect pairwise relationships across these dimensions. In the Novice, all the wines are clustered by white or red and all wines within each cluster are relatively similar. In the Sommelier, the wines are clearly separated, the two light reds are close together, the rosé sits alone, and the white and orange are relatively similar, yet very distinct from the reds. In this sense, we can say the sommelier has a more articulated structure, with finer local discriminations. When mapping the sommelier to the novice (**C**) the area in which the sommelier's experience can map down by adjunction onto the novice is relatively small. The sommelier loses some complexity, but its downward-mapping is so heavily constrained by the domain of the novice that the points have a clear, principled and lossy relationship. However, in the reverse (**D**), the possible domain of wine-experience in the sommelier is far larger than the novice, meaning that any given wine (and its corresponding arrow) in the novice has a far greater range of possible positions in the structure of the sommelier.

Consider then two quite different minds, such as my own and that of a bat, which, given there are great differences in anatomy and behaviour between us, likely share very few relations in our experience of a tree (Figure 3C). We cannot begin as the Meroitic linguist does, because a bat cannot report its experience to us (however, behavioural measures that recover mathematical structures are in principle possible in non-human animals, see (Nakayama et al., 2022)). If Qstr's inter-phenomenal route is in practice closed, we must unfold the bat's structure intrinsically, from its substrate. Assuming I already possess a Rosetta Stone sorted by modality, then I have validated which of my substructures specify the experience of visual extended space (undirected 2D grids in V1/V2 (Haun and Tononi, 2019)) and which specify auditory temporal experience (1D grids in A1/A2 (Comolatti et al., 2025)). Therefore, I can ask which of my modalities most resembles the unfolded cause-effect (sub)structure from the bat. A bat plausibly has some *kind* of experience of spatial extendedness, but the topology of its echolocation space likely differs considerably from mine. Provided that the substrate of the bat's experience still shows some kind of undirected 2D grid motif (thus specifying a corresponding space cause-effect substructure), I could perhaps say only that it has a space-*like* experience in the broad sense (Figure 2C). Should it actually align more closely with my auditory substructure than my visual one, I have principled grounds to infer that echolocation feels, for the bat, more like *my* hearing than seeing even while the narrow qualia of returning echoes from ultrasonic squeaks remains beyond me (a case of miscommunication described in Box 1). This is because we would establish a shared *grammar* of experience, similar to how we can uncover the grammar of the Meroitic language by performing an adjunction between two structures of words. The possible inferences here would be relatively epistemically unconstrained, especially compared to those we suggest in §5.1 and §5.2.1, but that is not to say they are without value. Being able to make a *principled* inference as to the experience of another mind (especially say, that of a non-human animal) is a marked advance on the inferences possible absent such structural approaches (Ellia and Tsuchiya, 2026).

## 5.3 No Relations

At some point, two minds will share so few relations that there can be no alignment between any experiential structure corresponding to their respective experiences. Consider encountering an alien. Without intelligible reports which is even more likely than the case outlined in §5.2, we may be unable to recover its qualia structure at all. However, should we be able to unfold a cause-effect structure from its substrate, we can say, on IIT, that it has *a* conscious experience. However, if nothing in that cause-effect structure aligns to anything within my own, my inference simply runs out (Figure 2D). With no shared relations to anchor my structure to its structure, the space of possible inferences is highly unconstrained. This limit is still principled in that the structures *themselves* tell me that no inference is licensed rather than a failure to access it at all. To run the metaphor of the Rosetta Stone to its end, a script for which no bilingual inscription exists, and in which no shared grammar can be found, cannot be translated at all.

# 6. Validating the Construction of a Rosetta Stone

Each claim we make above (§5) presupposes a valid Rosetta stone. So how can we be sure of having one in the first place? Any account of experience must be validated on ourselves (who are uncontroversially capable of conscious experience) first, and then be extended to the more complex cases we present here. Given the resulting mathematical structures of Qstr and IIT differ in provenance, they have different paths to validation which we will briefly make mention of here.

IIT's aims to account for both the presence and quality of consciousness. Regarding presence, the requirement (postulates) that a substrate have a large repertoire of alternative states and that these states be highly integrated has led to efforts to test IIT through perturbation of the human brain (Massimini et al., 2005). Delivering transcranial magnetic stimulation (TMS) to the cortex engages near-deterministic interactions among groups of neurons and the complexity of their response can be measured as a proxy for integrated information (the perturbational complexity index (PCI)) (Casali et al., 2013). Validated against subjects whose consciousness was confirmed by report, PCI discriminates conscious subjects from unconscious ones and identifies unresponsive patients who may retain a capacity for consciousness (Casali et al., 2013; Casarotto et al., 2016). As PCI was derived from IIT's postulates it has construct validity, not just empirical validity (Tononi and Boly, 2025). Regarding quality, IIT's explanatory identity conjectures that a property of experience corresponds to a property of the cause-effect structure (§4), with nothing left over. Therefore, IIT predicts that changing the structure (via the substrate that specifies it) should result in a corresponding change in experience, and vice-versa, even if activity is unchanged. For example, transient changes in the connectivity of undirected 2D grids via Hebbian learning leads to corresponding perceptual contraction of visual space (Song et al., 2017). Lesions of grid-like neurons in the visual cortex do not blind a hemifield, rather the hemifield ceases to exist where patients are typically unaware that something is amiss at all (Haun and Tononi, 2019; Corcoran et al., 2026). Even the counterintuitive prediction that a largely inactive cortex should still specify spatial extendedness is tentatively supported by meditators experiencing luminous extension in 'pure presence' (Boly et al., 2024). Further efforts to validate the link between structure and experience are also underway, see (Melloni et al., 2024; Takahashi et al., 2025). Similar predictions and tests are under development for the account of temporal experience (Comolatti et al., 2025). Each intervention adds a line to my Rosetta Stone, validating which part of the structure corresponds to a part of my experience. As evidence for this explanatory identity accumulates, it can be used in the future to reason from structure to phenomenology for contents of experience that introspection cannot adequately reach, such as pain, in the way that has already been done for spatial extendedness (Haun and Tononi, 2019) and time (Comolatti et al., 2025).[15]

Because qualia structures are recovered from experience, in order to validate Qstr we must ask whether these structures actually track phenomenology rather than just report. For example, withdrawing spatial attention collapses the qualia structure of shapes, but not faces (Rowe et al., 2025); structures can be recovered without verbal labels in the peripheral visual field (Zeleznikow-Johnston et al., 2023); children, who often confuse the labels of colours, have highly similar dissimilarity structures to adults (Moriguchi et al., 2025); building a structure from 'preference' rather than 'dissimilarity' tests whether the recovered structure is an artefact of the relation measure used (Matsuyanagi et al., 2025) and finally, similarity structures obtained from invisible stimuli do not align with conscious qualia structures in humans (Qianchen et al, in prep).

[15] An intriguing question is whether the number of experiential kinds is limited, because the kinds of substrates that can support them are limited. For example, the IIT formalism could imply that only certain architectures (or combinations thereof) can constitute a substrate of consciousness. It may turn out that, besides undirected grids, directed grids, rooted trees, and cliques (notwithstanding their countless variations), not so many kinds of substrates can do so. This would imply that the kinds of experiential contents are just as limited, and could be roughly categorized as belonging to extensions (space, time, and conceptual hierarchy) or local qualities (narrow qualia like hue and timbre), whose specific character could still be unbeknownst to us.

# 7. Conclusion

If we were handed a mathematical structure, how would we know what experience it corresponds to? To answer this problem, one seeks a Rosetta Stone, that is, a principled method to translate structure (a *description*) into experience (an *acquaintance*) (Chalmers, 2012, 2023). If *the having is the knowing*, this seems an impossible task. However, as we have argued, we do not find ourselves in such a position if the structure is built the other way around—starting the translation from experience and deriving the corresponding mathematical structure: a *description* built from *acquaintance*, one in which the explanatory identity between the phenomenal and the causal can be validated. Validated across many experiences and many subjects, this Rosetta Stone can be used to make inferences about other minds, serving as a dictionary to compare different experiences—much as we compare different languages, where the meaning of a word can be translatable with varying degrees of accuracy. This is because mathematical structures are a common formal medium through which different subjective experiences can be compared. The degree to which the mathematical structure of another mind corresponds with *my own*, determines how constrained the inference is that I may make. If the qualia structure, substrate anatomy and cause-effect structure each align, that inference is richer. If only some of these constraints are available, the inference is much coarser. Whenever I compare my structure to another's, in each case I construct something between an imperfect dictionary and a grammar; a system of principled constraints on what may be inferred given that I cannot truly be acquainted with another mind without becoming it.

These are not hypothetical cases either, alignment of *inter-phenomenal* qualia structures recovered from reports across (Zeleznikow-Johnston et al., 2023; Kawakita et al., 2025) and within (Togashi et al., 2026) individuals using GWOT is aiming to realise the necessary condition set out in §3. Intra-phenomenal alignment between cause-effect structures to supply sufficiency is being actively pursued using the same method (Togashi et al., in prep). Moreover, the two sides are converging as Qstr turns to align information structures (as proxies of cause-effect structures) recovered from brain data (Oizumi et al., 2025; Takeda et al., 2025; Maier and Tsuchiya, 2026). For all of these efforts, we establish a conceptual backdrop to infer the meaning of such alignments.

Finally, consider the alternative; that the question of other minds remains the entirely intractable epistemic problem it has been since one human first asked what it truly *is like* to be another. A grammar of another mind, with an imperfect dictionary at its best, is far more than we have ever been principled in claiming before.

**Box 1: Communicating meaning requires similar cause-effect structures.**

Shannon information theory, a mathematical theory of communication, is concerned with the probability of messages (in the form of symbols) and not with their meaning. IIT contends that *meaning* is not carried by symbols but *is* a structure (the interlocutor's cause-effect structure unfolded from their substrate in a state corresponding to their experience) (Zaeemzadeh and Tononi, 2024). The content of a message is given by the "shape" of the cause-effect substructure (Φ-fold) the message triggers. Only a substrate that specifies such a structure can bear meaning at all.

If meaning is accounted for by intrinsic structure, then to successfully communicate content between minds the sender must use symbols (in the Shannon information sense) to trigger a structurally-similar Φ-fold in the recipient, disassociating meaning from information processing. This has three down-stream consequences. First, communication of Shannon information can be perfect, such as between computers, yet if the substrate does not specify a cause-effect structure no meaning is communicated at all. Second, if the communication of Shannon information is blocked between sender and receiver where both are conscious high-Φ systems, the sender's intended Φ-fold cannot be triggered in the recipient. Finally, Shannon information can be high and both parties support cause-effect structures, yet the sender's intended Φ-fold will not be triggered in the recipient if both parties have substantially different substrate architectures.

Successful communication between humans that conserve's structure, such that both parties arrive at the same meaning can be thought of as a strong adjunction like we present here. Assuming the human reader of this chapter understands English, understanding our argument should correspond to symbols (words) triggering successive Φ-fold's that have a strong adjunction to the authors'. Translation weakens this adjunction without breaking it. A good translation replaces the sender's symbols with symbols that can trigger a Φ-fold in the receiver as similar as possible to the sender's, but because words in different languages may have different connotations, the meaning recovered is likely similar rather than the same. Note that Shannon information cannot register this: over a high-fidelity channel the mutual information between the sent symbols and the received symbols is maximal whether or not the sender's Φ-fold can trigger a similar one in the receiver. Weak adjunction would occur where both parties have different substrate architectures, such as a human and a dog. Here the sender may not entirely trigger the intended Φ-fold in the recipient. Consider when I mention the word 'walk' around my dog, who goes from resting to excitedly wagging its tail. My dog does not understand English, but may recognise this specific word such that a corresponding Φ-fold is triggered in his cause-effect structure and some meaning is communicated. Here even less meaning survives than under translation, since the dog's substrate specifies very different distinctions and relations from mine, even though the acoustic signal reaches him undiminished.

# Acknowledgements

We would like to thank Prof. Michael Pauen for his help in editing an early version of this manuscript. NT was supported by National Health Medical Research Council (GNT2037172), Australian Research Council (DP240102680), Japan Society for the Promotion of Science Grant-in-Aid for Transformative Research Areas (A) (23H04829, 23H04830) and Japan Science and Technology (JST) Moonshot R&D Grant (JPMJMS2295-14), Theoretical Sciences Visiting Program, Okinawa Institute of Science and Technology.